\documentclass[letterpaper]{article} 
\usepackage{aaai24}
\usepackage{times}  
\usepackage{helvet}  
\usepackage{courier}  
\usepackage[hyphens]{url}  
\usepackage{graphicx} 
\usepackage{natbib}  
\usepackage{caption} 
\usepackage{algorithm}
\usepackage{algorithmic}

\usepackage{newfloat}
\usepackage{listings}
\DeclareCaptionStyle{ruled}{labelfont=normalfont,labelsep=colon,strut=off} 
\floatstyle{ruled}
\newfloat{listing}{tb}{lst}{}
\floatname{listing}{Listing}
\usepackage{multirow}
\usepackage{xcolor}

\nocopyright

\title{Lightweight Knowledge Representations for Automating Data Analysis}
\author{
    Marko Sterbentz,
    Cameron Barrie,
    Donna Hooshmand,
    Shubham Shahi,
    Abhratanu Dutta,
    Harper Pack,
    Andong Li Zhao,
    Andrew Paley,
    Alexander Einarsson,
    Kristian Hammond
}
\affiliations{
    
    Northwestern University\

    Computer Science Department\\
    Evanston, Illinois\\
}

\begin{document}

\maketitle

\begin{abstract}
The principal goal of data science is to derive meaningful information from data. To do this, data scientists develop a space of analytic possibilities and from it reach their information goals by using their knowledge of the domain, the available data, the operations that can be performed on those data, the algorithms/models that are fed the data, and how all of these facets interweave. In this work, we take the first steps towards automating a key aspect of the data science pipeline: data analysis. We present an extensible taxonomy of data analytic operations that scopes across domains and data, as well as a method for codifying domain-specific knowledge that links this analytics taxonomy to actual data. We validate the functionality of our analytics taxonomy by implementing a system that leverages it, alongside domain labelings for 8 distinct domains, to automatically generate a space of answerable questions and associated analytic plans. In this way, we produce information spaces over data that enable complex analyses and search over this data and pave the way for fully automated data analysis.

\end{abstract}

\section{Introduction}

\begin{figure*}[ht]
\centering 
\includegraphics[width=\linewidth,keepaspectratio]{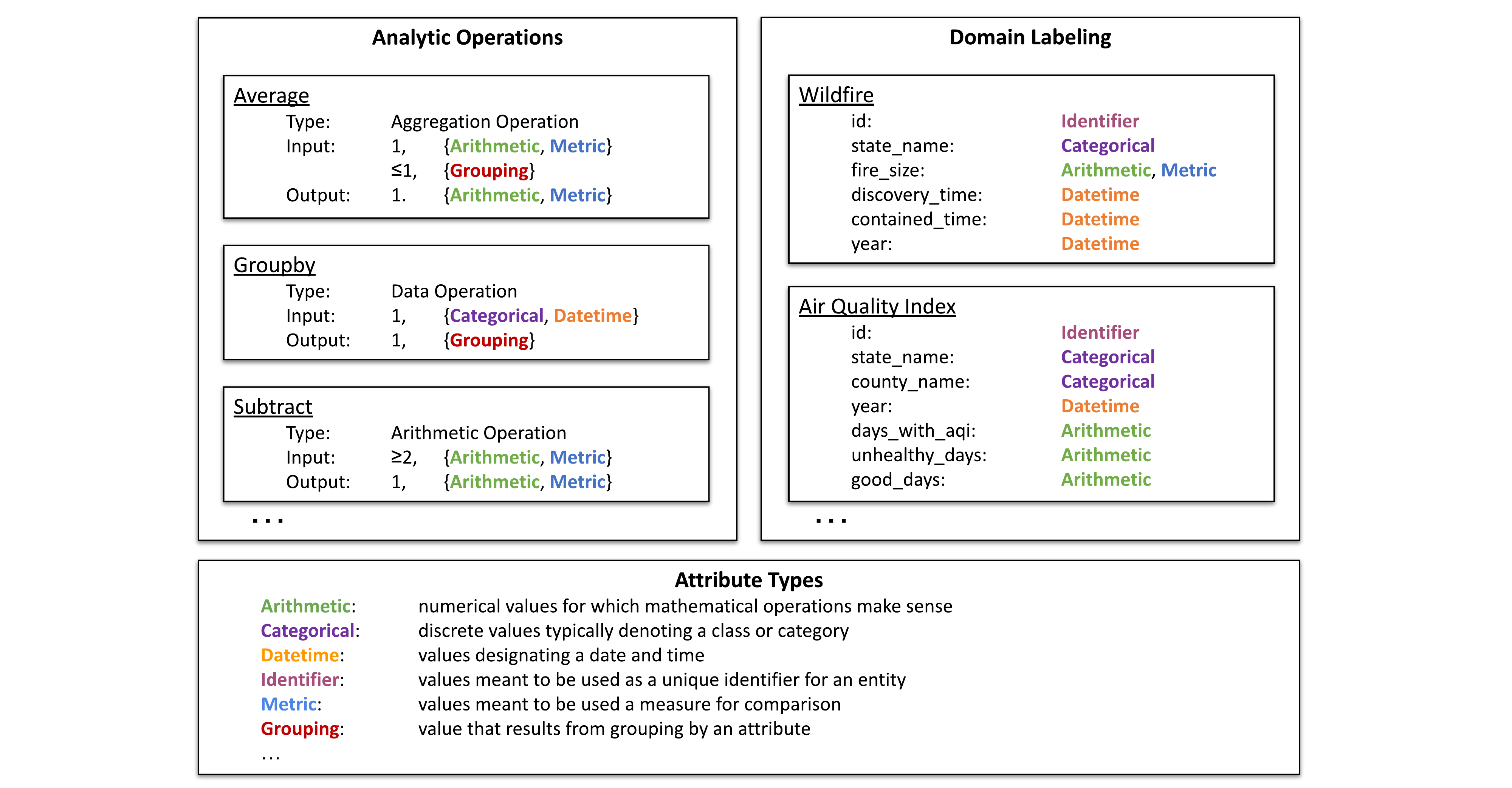}
\caption{In the top left box, analytic operations as defined by the taxonomy we propose. Each operation has a type which specifies an "isa" relation with its parent type. The inputs and outputs are defined according to the number of inputs and the \emph{attribute} types which can be passed in. For example, the Subtract operation shown above takes in two or more inputs which can have a type of either \emph{Arithmetic} or \emph{Metric}, and it will output one value which will have types of \emph{Arithmetic} and \emph{Metric}. The meanings of these attribute types are shown in the bottom box. In the top right are two \emph{entities} from a single domain labeling. For simplicity, only the \emph{attributes} and their types are shown. The types for each attribute provide an indication of how these \emph{attributes} should be analyzed by operations in the analytic taxonomy to produce meaningful information. More in depth discussion of these concepts are provided in Section \ref{sec:methods}.} 
\label{fig:analytic_taxonomy}
\end{figure*}

Data science is a vast and interdisciplinary field whose principal goal is to derive meaningful information from data. The data science pipeline traditionally consists of data ingestion, cleaning this data, analyzing the data, and then presenting results to stakeholders \cite{biswas2022art}. In this work, we focus solely on automating the data analysis portion of this pipeline. To carry out data analysis, data scientists use domain knowledge, the available data, the operations and algorithms that can be used to derive information from that data, and how all of these facets interweave to develop a space of analytic possibilities and reach their information goals. Due to the technical skills required, performing good data analysis is a very demanding task, and it takes a significant amount of training to do it well.

With the vast amounts of data being generated on a daily basis, it is becoming increasingly difficult for organizations to produce high quality and accurate information from it all. This is particularly problematic in domains with high stakes such as healthcare, journalism, and policy making since access to high quality and accurate information is crucial for sound decision making. However, while data scientists are employed across a wide range of such domains, it is not feasible for every organization to hire the requisite number of data scientists due to both cost constraints as well as an ever increasing demand for data scientists. 

Furthermore, not every domain expert has the technical skills needed to translate the questions they want answered into queries and analysis against the available data. As a result, data scientists typically serve as the bridge between these domain experts and the data. Data scientists work with domain experts who have extensive knowledge of the data and the real-world entities it describes, and then apply their knowledge of analytic operations and algorithms to produce the desired information. Importantly, there is a clear separation of concerns between domain knowledge and analytic knowledge. We will take advantage of this when producing the lightweight knowledge representations that will comprise the foundation of our approach for automating data analysis.

Within the current literature, there are a variety of approaches to automating data analysis. AutoML libraries like Auto-WEKA \cite{thornton2013auto}, auto-sklearn \cite{feurer2015efficient}, and AutoKeras \cite{jin2019auto} seek to automate the model building process, but still require a data scientist to make decisions about how best to utilize the data. The method that comes closest to enabling non-technical experts to derive information from data is \cite{paley2021data}, in which the authors sought to enable domain experts to more easily perform data analysis via a notebook style interface. However, their representations of analytic operations and domain knowledge are lacking. The set of analytic operations they implement specify the data they can operate on based on native relational database types (i.e. integer, varchar, boolean). This will ensure that the analysis being performed is structurally valid. However, this does not mean that the results will be meaningful as well. That is, knowing that an average \textit{can} be applied to any numeric column is far less useful than knowing that it \textit{should} be applied to columns which represent the key metrics which will be used for evaluation. Additionally, existing operations cannot be composed into more complex operations. This is due to the lack of typing information for the outputs of the operations, as well as the rigid plan representations that requires a single analysis operation to be performed on a single data column with optional grouping and filtering. These issues are all problematic on their own, but taken together they severely limit the scalability of this approach.

Our primary goal is to enable autonomous systems to be able to make the same kinds of decisions that data scientists make to derive \textit{meaningful} information from data. In this paper, we lay the foundations for automating data analysis by building representations for the internal knowledge and processes a data scientist utilizes when deriving information from data. We present a novel domain-independent and expandable taxonomy of analytic operations that captures knowledge needed to automate data analysis. In order to know how these operations should map to available data, we define a taxonomy of \textit{attribute types}. The inputs and outputs of the analytic operations are defined in terms of the \textit{attribute types} which both constrain the analysis to produce useful information and enable operations to be easily composed into more complex operations. We also present a formal \emph{domain labeling} which specifies complementary domain knowledge. With this \emph{domain labeling}, the tables and columns of a relational database are mapped to entities, attributes, and the relationships between them. Each attribute maps to an underlying column of the data and is given an \textit{attribute type} that links the data to the analytics taxonomy. Since the creation of a domain labeling is done manually, we put a particular emphasis on ensuring that the domain-level expansion process is simple, with the aim of enabling domain experts to bring in data and explore the space of information it encodes. A sample of these knowledge representations is provided in Figure \ref{fig:analytic_taxonomy}.

We also develop an analytic plan representation that enables complex analyses to be composed in terms of the entities and attributes present in the \textit{domain labeling}. This representation is agnostic with respect to the underlying data medium. This means that details like table joining, which are often a part of SQL queries, are abstracted away. The hope is this representation is easier for data scientists to write, as well as more interpretable for those without a data science background than query languages like SQL and SPARQL. We build an analytics engine which is capable of executing these analytic plans by generating the required queries against the underlying data.

With this work, we seek to bring out the space of meaningful information based on the available data and analytics. To this end, we develop a planning component that can generate a wide variety of useful analytic plans for a given domain labeling. We build a language generation component which can produce the question this analytic plan is meant to answer. In this way, the space of analytic possibilities can be surfaced in natural language. To demonstrate the generality of our approach, we create domain labelings for 8 different domains: healthcare, urban housing, criminal justice, environmental sustainability, education, legal and judicial, socioeconomics, and business. We use these in conjunction with our planner to showcase the kinds of useful information our approach can present to a user. The ability to generate this plan space paves the way for the future development of automated search techniques and greater automation of data analysis.

The primary contributions of this paper are:

\begin{itemize}
    \item A lightweight and extensible taxonomy of data analytic operations that scopes across domains and data.
    \item A method for producing domain-specific labels that links this analytics taxonomy to actual data and provides linguistic constructs that enable improved communication of any derived information.
    \item An execution agnostic plan representation that enables the retrieval and analysis of data according to the entities in the domain labeling.
    \item A method for leveraging the analytics taxonomy and domain labeling in order to build searchable information spaces over data that surfaces meaningful information in natural language, making complex analyses and search over this data easily available to non-technical human experts.
\end{itemize}

\section{Related Work}

\subsubsection{Existing Upper Ontologies}
Explicitly representing the knowledge that a data scientist uses can provide a means to automate the decisions data scientists must make in order to carry out data analysis. Formalizing world knowledge with structured representations in the form of an ontology has long been a way to imbue systems with an understanding of the entities and features of the domain they operate in. Such knowledge representations provide the structure necessary for a wide range of reasoning and inference processes to function. Ontologies such as Wikidata \cite{vrandevcic2014wikidata} seek to codify general world knowledge, ontologies such as Cyc \cite{lenat1995cyc} are designed to encode commonsense rules as well, and ontologies such as Basic Formal Ontology (BFO) \cite{arp2015building} aim to promote integrability among domain ontologies empirically. Existing ontologies and representational languages provide some degree of mathematical knowledge or ways to incorporate it \cite{lange2013ontologies, angles2019rdf}. However, their primary focus is to provide a means for automated proof solving rather than data analysis. Furthermore, while large scale ontologies such as these are specified enough to be used for a diverse range of areas, they are cumbersome to extend for anyone not already steeped in complex web of relations they specify. This limits their utility when the expected user is a domain expert who has never been exposed to such formalisms. This makes them poor candidates for a representational medium for our purposes.

\subsubsection{Domain-Specific Knowledge Representations} 
It is often the case that highly specialized ontologies are developed for use within a particular domain. Indeed, such knowledge representations have been developed for a diverse range of areas such as medicine \cite{salvadores2013bioportal}, law \cite{casellas2011legal}, food \cite{kamel2015towards}, chemical engineering \cite{marquardt2010re}, and biological environments \cite{buttigieg2013environment}. However, the production of ontologies such as these requires extensive expertise in ontology design and substantial amounts of time since they require end-users to survey their respective domain in order to use them. This can be a challenging task for non-technical users. What would be more useful is a formal representation of data analytic knowledge that can scope across multiple domains with as little configuration required as possible.

\subsubsection{Knowledge Representations for Data Science}
The creation of knowledge representations for data science has recently emerged as an area of interest. For instance, \cite{patterson2019teaching} have semantically enriched data science scripts with the goal of successfully modeling computer programs. However, their work focuses more on supporting automated reasoning about data science software rather than encoding core analytic knowledge and processes that can be used when mapping analytics onto data in a domain-agnostic fashion. \cite{sicilia2018ontologies} presents a way to describe the data transformations that can be applied as part of a data science pipeline. Our work includes this as a subcomponent of a model for describing how analytics map onto data and how formalized domain knowledge can be utilized to determine the kinds of analyses that will result in meaningful information. In an effort to codify knowledge around data, as well as to increase interoperability and reusability of methodologies, multiple other representations have been created. Examples of these include DMOP for data mining processes \cite{KEET201543}, PMML for machine learning models \cite{guazzelli2009efficient}, SDMX \cite{capadisli2015linked} and STATO \cite{gonzalez2016statistics} for statistics, and even some ontologies for modeling data transformation and workflows \cite{bowers2004ontology} \cite{barker2007scientific}. 

\subsubsection{AutoML for Data Analysis}
There exist approaches that seek to make it easier to perform data analysis. AutoML libraries such as Auto-WEKA \cite{thornton2013auto}, auto-sklearn \cite{feurer2015efficient}, and AutoKeras \cite{jin2019auto} aim to make it simpler to train high quality models given a set of data. However, these are aimed at providing tools that speed up the workflow of established data scientists and require that the user make complex decisions about how data should be utilized within the scope of these tools. Automating these decisions would provide an avenue for expanding access to the high quality information which data scientists produce. Libraries such as these could ultimately be incorporated into a system based on the knowledge representations we present in the following section.

\section{Methodology}
\label{sec:methods}

\begin{figure*}[ht]
\centering 
\includegraphics[width=\linewidth,keepaspectratio]{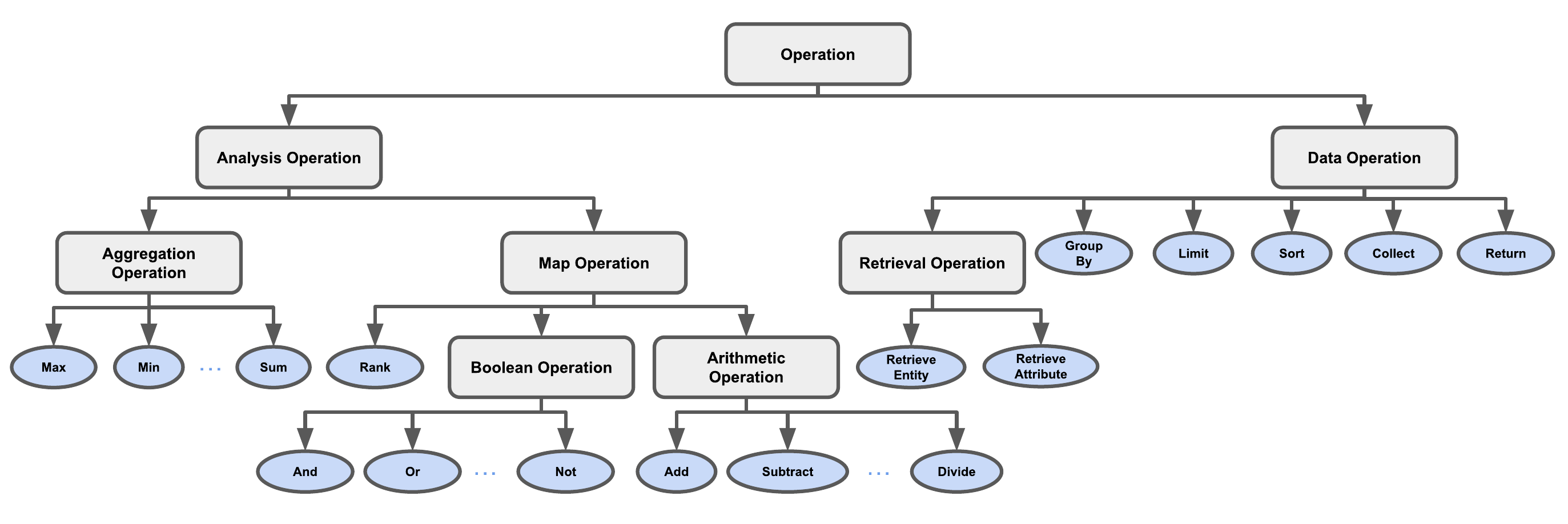}
\caption{The analytics taxonomy which encodes knowledge of data analytic operations and processes. It is designed to be extensible and, in addition to those shown in here, currently comprises 11 aggregation operations, 9 boolean operations, and 6 mathematical operations. This taxonomy specifies a conceptual mapping between these operations and the types of data that makes sense for them to operate upon, much in the same way a data scientist identifies the appropriate analysis to carry out on any given data.} 
\label{fig:operation_ontology} 
\end{figure*}

In this section, we present the data analytics taxonomy which provides a lightweight and extensible description of analytic operations and the kinds of data they are meant to operate on. We also present the domain labeling which is another lightweight representation built on top of data that provides information about the entities described by the data. Taken together, these two representations allow for analyses to be mapped to data such that information which is meaningful to an end user can be derived automatically.

\subsection{Data Analytics Taxonomy}

In order to automate the data analysis process, a representation of analytic operations, their functionality, and their inputs and outputs must be specified. Such a representation can be leveraged to determine what analysis can be performed on any given piece of data, how this data will be transformed as a result of this analysis, and how this result can be used as input to other analytic operations. The analytics taxonomy we present is shown in Figure \ref{fig:operation_ontology}.

\subsubsection{Analytic Operations}

Analytic operations take in attributes as inputs and produce an output which can be considered as a derived attribute. Not all analytic operations specified in the taxonomy can be or should be applied to all types of attributes. A key goal of the analytics taxonomy is to provide this mapping between analytic operations and the data. A simple way to do it would be to use the type information from the database (float, integer, varchar). However, naively applying an analytic operation, like an \emph{average}, to every integer or floating point attribute will result in useless information. In contrast, if we define attribute types for the data which indicate what an attribute is at a deeper level, then we can provide the necessary knowledge to an underlying analytic system so that it knows when it can effectively apply a piece of analysis in order to produce meaningful information. This is done by specifying an \emph{attribute type} for the inputs and outputs of each analytic operation. Examples of analytic operation definitions can be found in Appendix \ref{app:analytic_operations}.

\subsection{Attribute Types} We define six main attribute types which enable us to test functions against the data to determine what analytics can be performed. For example, it makes sense to perform a \emph{max} operation on a column of a patient's heart rate measurements, but performing a \emph{sum} operation on heart rate column serves no practical purpose. To provide the necessary information to determine when it makes sense to apply a piece of analysis, we specify type information for each attribute. The six main attribute types are: 

\begin{itemize}
    \item \textbf{Arithmetic}: numerical values for which mathematical operations make sense
    \item \textbf{Categorical}: discrete values typically denoting a class or category
    \item \textbf{Datetime}: values designating a date and time
    \item \textbf{Document}: values containing free text
    \item \textbf{Identifier}: values meant to be used as a unique identifier for an entity
    \item \textbf{Metric}: values meant to be used as a measure
\end{itemize}

By defining attribute types for the data indicating what the attribute is, the system can know when to apply a piece of analysis such that  meaningful/relevant information can be produced. For example, assume there is a dataset comprising business reviews with a \emph{Business} entity which has attributes \emph{number of reviews} and \emph{rating}. They would both be considered \emph{Arithmetic} attributes, while only \emph{rating} is a \emph{Metric}, as it can be used as a measure of the entity in the context of this dataset. These attribute types are leveraged to constrain the space of possible analytics that can be performed on the entities specified in a domain labeling.

We also define a set of attribute types reserved for \emph{derived attributes}, which are those attributes that result from executing an analytic operation. These include:
\begin{itemize}
    \item \textbf{Entity}: value that results from retrieving an entity
    \item \textbf{Attribute}: value that results from retrieving an attribute
    \item \textbf{AttributeCollection}: values that result from collecting many attributes
    \item \textbf{Group}: value that results from grouping by an attribute
    \item \textbf{Filter}: value that results from specifying a filter
    \item \textbf{Sort}: value that results from specifying a sorting order over the results
    \item \textbf{Limit}: value that results from specifying a limit over the results
    \item \textbf{String}: value that results from specifying a particular string to use
\end{itemize}

For each of the operations in the data analytics taxonomy, each of the input and output attributes are constrained to have one or more attribute types. For example, the \emph{average} operation would have exactly one mandatory input argument with an attribute type of either \emph{Arithmetic} or \emph{Metric}. It would have a second optional argument with the \emph{Group} derived attribute type. The output of this operation would be a single derived attribute with a type of either \emph{Arithmetic} or \emph{Metric}. For a full listing of the operations, operation type, and their input and output attribute types, please see Appendix \ref{app:operations}.

\subsection{Domain Labeling}

In order to effectively apply the operations specified in the analytics taxonomy to data, we produce a lightweight mapping called a \emph{domain labeling} which specifies the entities that comprise the data, their attributes, and the relationships between these entities. 
Each of the attributes has one or more attribute types which allows them to be linked to the analytics taxonomy. Importantly, these domain labelings are simple to specify relative to conventional domain ontologies and the process of creating them is more similar to tagging than the knowledge engineering that is applied when producing an ontology. In this section, we present further details of how this domain labeling is specified.

\subsubsection{Entities}

\begin{figure*}[ht]
\centering 
\includegraphics[width=\linewidth,keepaspectratio]{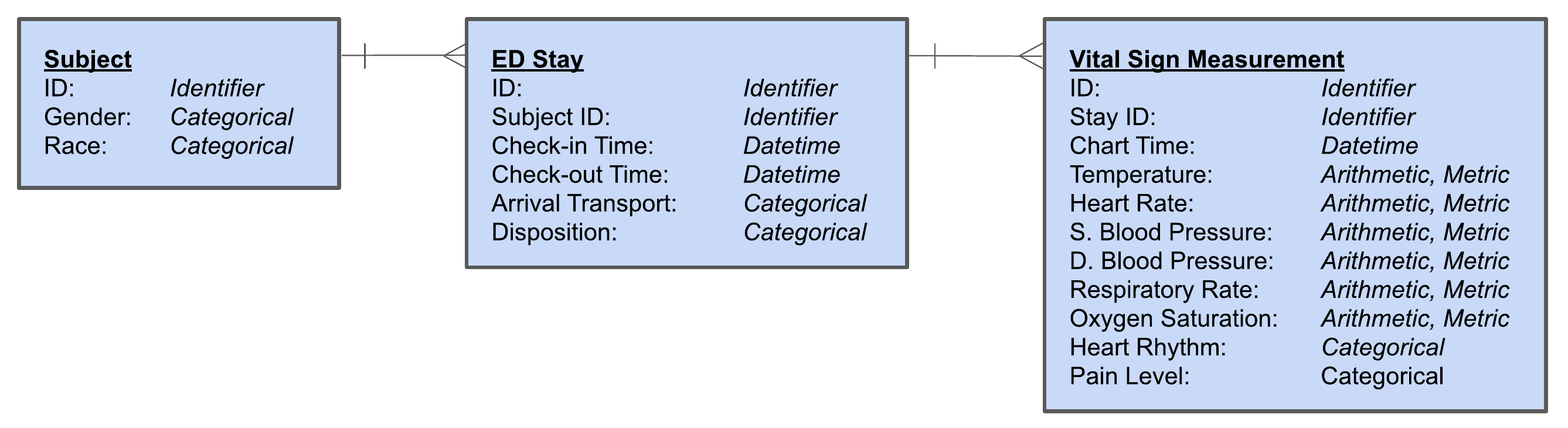}
\caption{A truncated example of a domain labeling which specifies \emph{entities}, their \emph{attributes}, and how they relate to each other for a dataset in the healthcare domain. The \emph{entity} name is given in bold, \emph{attributes} are listed on the left, and their corresponding \emph{attribute} types are shown on the right. Each of these \emph{attributes} maps to specific column of the underlying database, and each of the \emph{relationships} provides the joins between the underlying tables. This formulation enables analytic plans to be specified in terms of the \emph{entity} \emph{attributes} without regards to how the tables should be joined, thus abstracting out implementation details that can be automatically handled with an execution engine.} 
\label{fig:domain_labeling} 
\end{figure*}

\emph{Entities} defined by a \emph{domain labeling} specify one of the entities represented in the corresponding database. An \emph{entity} is a concept that groups relevant data together, potentially spanning multiple underlying data tables, into a single unified grouping. This \emph{entity} contains \emph{attributes} (which map to columns of the data that comprise the entity) and has \emph{relationships} with other \emph{entities} (which map to one or more joins across the data tables). Each \emph{entity} has a uniquely identifying name which is used to refer to it when performing analysis on its \emph{attributes}. An example of this can be seen for a subset of instances in the healthcare domain in Figure \ref{fig:domain_labeling}. For this dataset which details information about visits to the emergency room, entities include patients, stays, tests/diagnostics performed on the patient, and ER centers. Note that in Figure \ref{fig:domain_labeling}, for the domain depicted, only a subset of all the \emph{entities}, \emph{attributes}, and their types are shown. For examples of fully specified domain labelings, please see Appendix \ref{app:domain_labelings}.

\subsubsection{Attributes}

Each \emph{entity} has one or more \emph{attributes}, each of which maps to a column of one of the underlying database tables comprising the \emph{entity}. Each \emph{attribute} has a data type (e.g. integer, float, string) and one or more \emph{attribute types} (e.g. categorical, metric, arithmetic). The former is used when producing the object-relational mapping for the database and the latter is used to connect this \emph{attribute} to the operations from the analytics taxonomy that can be applied to this \emph{attribute}.

Each \emph{attribute} also has a "nicename" that provides a more descriptive label for the \emph{attribute} than the column name, as well as the units for this \emph{attribute}, if any. For example, the \emph{attribute} "dbp" could have the nicename of  "Diastolic Blood Pressure" and units "mmHg". These two properties of the \emph{attribute} are used by the language generator described in Section \ref{sec:question_gen}. 

\subsubsection{Relationships}

\emph{Relationships} define the connection between two \emph{entities} in the labeling. These \emph{relationships} can be one-to-one, one-to-many, or many-to-many. In the specification of the domain labeling, they are represented as an abstraction of one or more SQL joins. Adding \emph{Relationships} between \emph{Entities} forms a \textit{domain graph schema}, and each provides relevant metadata (e.g. one \emph{Subject} can have many \emph{Emergency Department Stays}). For instance, with a one-to-many \emph{relationship}, the \emph{entity} on the many side can be grouped by the \emph{entity} on the one side. This allows for aggregations to be applied to the \emph{entity} on the \emph{one} side. For example, the system knows it is possible to compute the average of stay duration grouped by subject ID based on the type of \emph{relationship} between the two \emph{entities} these \emph{attributes} come from.

\section{Analytic Plan Representation}

\begin{figure}[!ht] 
\centering 
\includegraphics[width=\columnwidth]{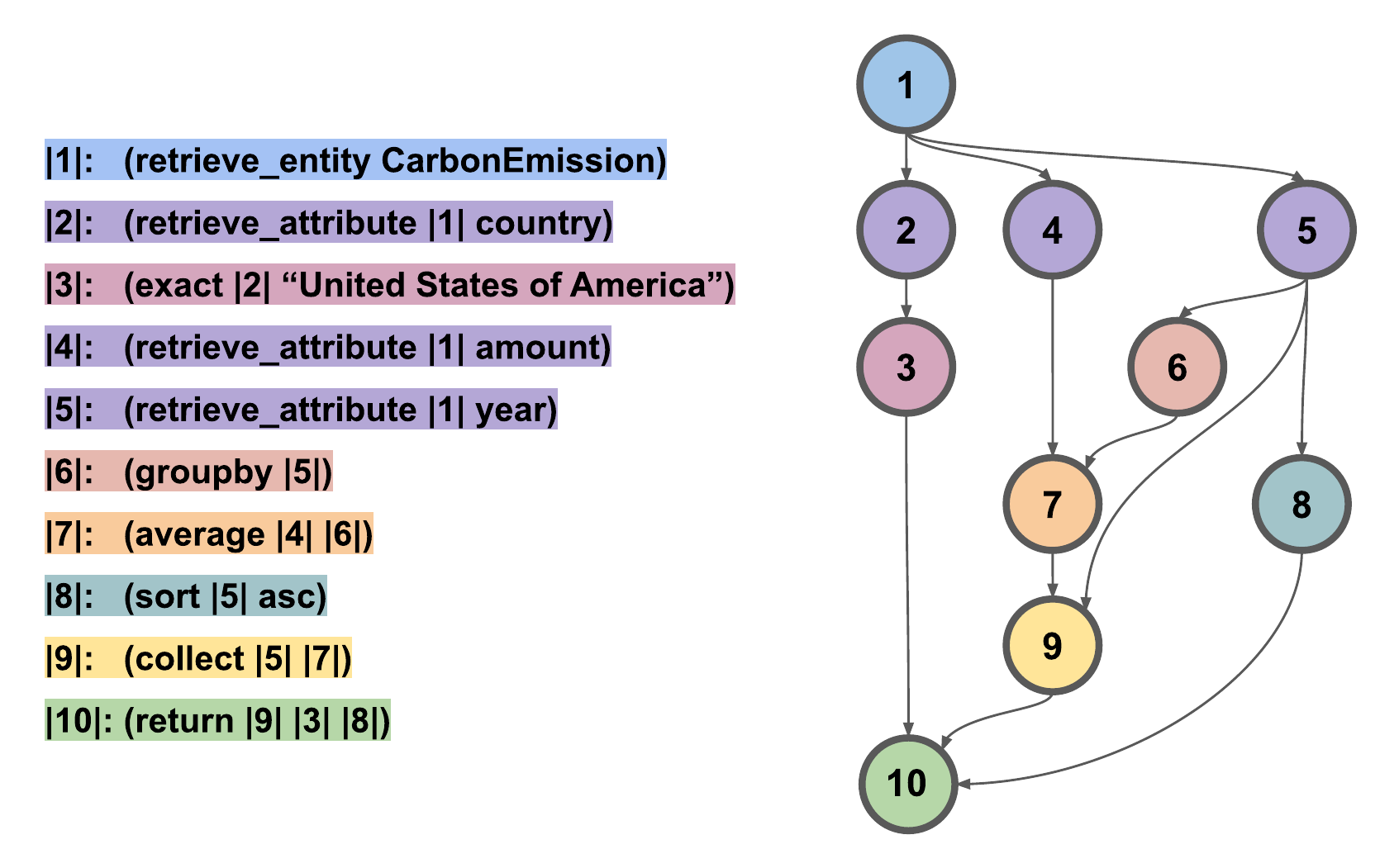}
\caption{The analytic plan in textual and graph forms are shown for the question: “What is the average carbon emissions grouped by year in ascending order for the United States of America?”} 
\label{fig:plan_example} 
\end{figure}

In order to effectively use the domain labeling and analytic taxonomy, we require an expressive and compositional plan representation. Existing representations like SQL are inadequate since the analytics taxonomy scopes beyond the operations supported within this query language. Additionally, this higher-lever plan representation abstracts away specific implementation details of the specific underlying query language (e.g., SQL joins), meaning not only is it simpler to use, but also that it is agnostic to the data storage format and corresponding query language. This allows underlying implementations to seamlessly expand beyond SQL operations in the future (e.g., to ontological or textual data). To satisfy this representational need, we define an analytic plan representation which allows for the specification of plans in which the entities and attributes defined in the domain labeling are retrieved and analyzed using the operations defined by the analytics taxonomy. An example of this plan representation can be seen in Figure \ref{fig:plan_example}.

\subsection{Plan Structure}
Plans are represented as a directed acyclic graph that specifies an ordered series of steps to carry out, wherein operations are chained together in order to retrieve and analyze data. Each node of the graph represents an operation whose output is fed to later steps that require the results. In this way, arbitrarily complex plans can be composed to satisfy any information goal that can be described with the available data and analytics.

Leveraging the analytics taxonomy allows this representation to support standard operations in the underlying query language (e.g., SQL), for example, retrieval, aggregation, groupby, filtering, sorting, etc. \emph{Entities} and \emph{attributes} described by the domain labeling can be retrieved using the \emph{retrieve\_entity} and \emph{retrieve\_attribute} operations. Analytic operations take these attributes as input and produce \emph{derived attributes} (i.e. ones which are not present in the domain labeling), which can be used as input to subsequent operations. In this way, arbitrarily complex analysis can be composed by chaining operations together. Data manipulation operations such as \emph{sort} and \emph{limit} are also supported, as are filtering operations like comparison and boolean operators.

Both entity attributes and derived attributes can be passed to the \emph{collect} operation in order to stage them for output in the final results. The \emph{return} operation takes in the attributes to be collected, along with any \emph{sort}, \emph{filter}, and \emph{limit} operations that were specified. Each \emph{return} operation denotes the end of a plan and results in a structure which is analogous to a single SQL query. Subsequent queries can retrieve \emph{attributes} collected by prior queries. In this way, subplans, which are analogous to nested queries or "subqueries" found in SQL, can be represented.

\section{Analytics Engine}
Execution of analytic plans requires that they first be converted to a query format that is native to the datasource (e.g., SQL for relational databases) and then executed to retrieve results. This is where the analytics engine comes in.

\subsection{Plan Parsing and Execution}
Conversion of the graph-structured analytics plan into a query language is done by first breaking the graph into "subplans", where the result of one subplan functions as a data source for subsequent subplans. From each subplan, the necessary information to form an executable query, including \emph{entities}, their \emph{attributes}, analytics operations, and filters, is then identified. For example, in the plan in Figure \ref{fig:plan_example}, the attributes \textit{country}, \textit{amount}, and \textit{year} are retrieved from the \textit{CarbonEmission} entity. The \textit{average} operation is applied to \textit{amount}, grouped by \textit{year}. Lastly, the results are sorted by \textit{year} and restricted such that \textit{country} has value "\textit{United States of America}". From this information, the query is constructed using a query abstraction library (for relational databases we use the SQLAlchemy Python package \cite{sqlalchemy}). In this last step, the domain labeling is leveraged to convert the \emph{entity} and \emph{relationship} abstractions to the proper tables and joins.

\subsection{SQL Object Relational Mapping}
While the system’s analytics engine is designed to be extendable to a variety of data source types, it is currently only configured to execute queries against relational databases. Upon initialization, for each selected domain labeling, a corresponding object relational mapping (ORM) is built using SQLAlchemy. The ORM provides a programmatic interface between the information defined in the domain labeling, and the data stored in a relational database.

The ORM is constructed using configuration mappings defined in the domain labeling, specifically, the tables, columns, and joins between tables. Unlike objects defined in the domain labeling, the ORM objects hold a direct one-to-one correspondence with database objects; ORM entities correspond to tables, attributes correspond to columns, etc.

\subsection{Implicit Joins}
A major benefit provided by the abstractions of the domain labeling is that, for a given domain, relationships between tables need only be defined once (in the configuration mapping of the domain labeling). No join information of any kind is required in plan definitions. Instead, the system leverages the joins and \emph{relationships} defined in the domain labeling to determine which SQL joins to use when \emph{attributes} are selected corresponding to columns of different tables.

First, all necessary joins between tables within an \emph{entity} are identified. These intra-entity joins are necessary when multiple \emph{attributes} are specified as belonging to the same \emph{entity}, but correspond to columns from different tables. Then, all joins between each pair of \emph{entities} in a given plan are identified by collecting joins along the shortest path of \emph{relationship} links between those \emph{entities}.

\section{Evaluation}

\begin{figure}[!ht] 
\centering 
\includegraphics[width=\columnwidth]{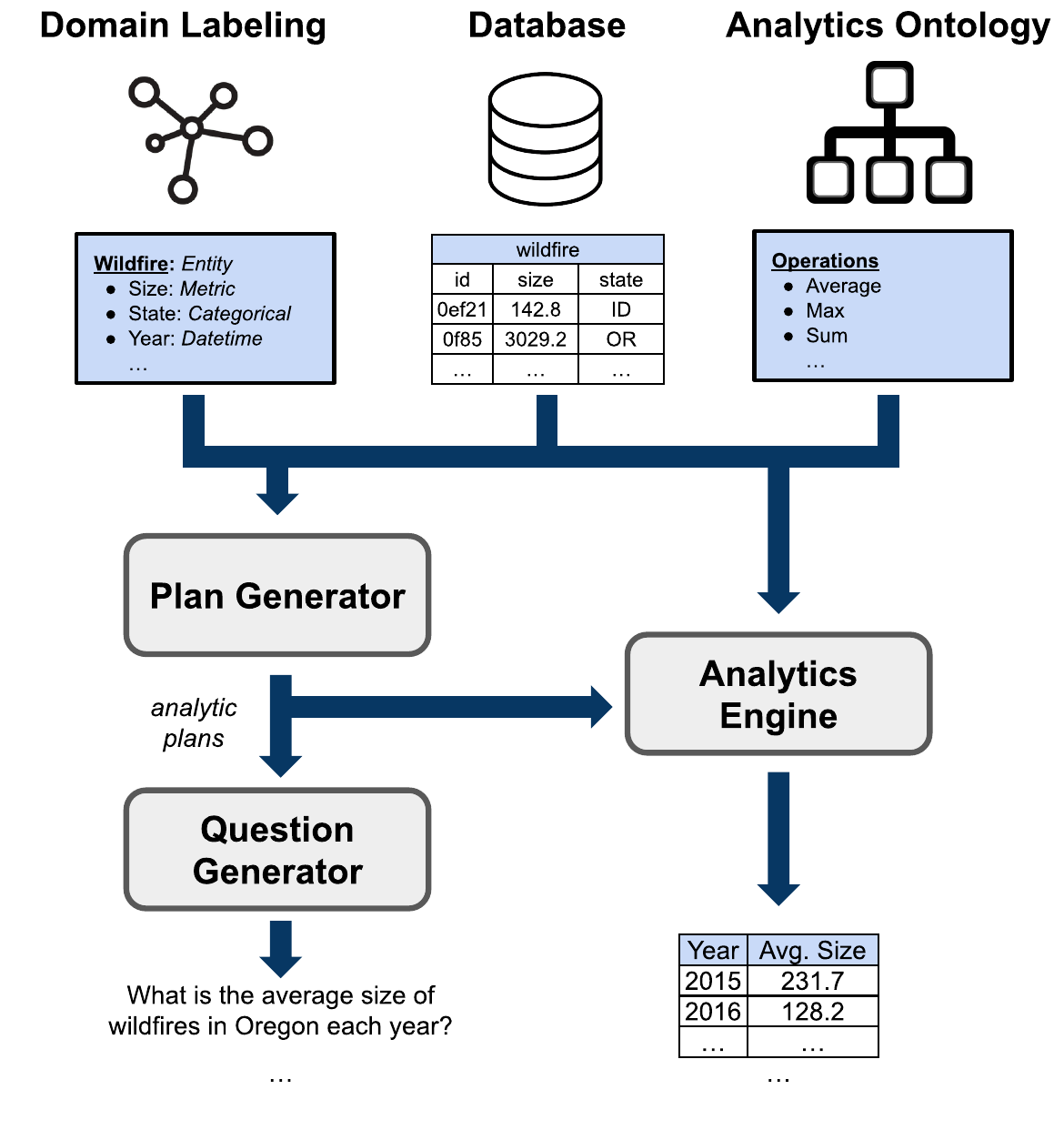}
\caption{The domain labeling encodes knowledge of the \emph{entities} present in a database, while the analytics taxonomy encodes knowledge of the core analytic processes which provide a conceptual framework for performing analytics on these \emph{entities}. This knowledge is passed to the plan generator, which constrains its generation of plans to a subset of meaningful plans based on the \emph{attribute} types. For each of these plans, the question generator produces a natural language question. By surfacing these plans in natural language, we provide a means by which people can easily search the space of meaningful information, and pave the way for more automated methods, such as by using vector embedding models.} 
\label{fig:ontology_implementation} 
\end{figure}

To illustrate how the analytics taxonomy can be operationalized as a real system, we develop a prototype system that utilizes the operation taxonomy and a domain labeling to produce a set of useful analyses and information that can be derived from data described by the domain labeling. These analyses are presented as a set of natural language questions that can be answered using the data and available analytics. Each question has a corresponding analytic plan that specifies how to derive these answers from data, and the analytics engine is equipped with a code implementations of the operations specified by the domain-independent analytic taxonomy. The key idea with this approach is that anyone, including domain experts without data analysis skills, can search through the list of questions and pick ones they want the answer to. Figure \ref{fig:ontology_implementation} shows the process flow for exposing these information spaces to a user.

\subsection{Domains}

To demonstrate a variety of applications of our approach in a wide range of domains, we evaluate our system using data from 8 domains. We specify a domain labeling for each of these domains. Each labeling determines the domain-relevant features of the data and how they map to real world concepts and other in-domain \emph{entities}. Below is a list of domains and datasets for which we seek to surface the space of meaningful information possibilities.

\subsubsection{Education}
The Illinois Report Card is an annual report released by the Illinois State Board of Education that shows how the state, and each school and district, are progressing on a wide range of educational goals \cite{jmlarkin-2020}. The Report Card offers a complete picture of student and school performance in order to inform and empower families and communities as they support their local schools.

\subsubsection{Criminal Justice}
The Center for Homeland Defense and Security’s Shooting Incidents at K-12 Schools dataset \cite{center-for-homeland-defense-and-security-2023} describes shooting incidents based on publicly available data on such happenings from the beginning of 1970 through June of 2022. An incident is defined as any occasion when a gun is brandished, is fired, or a bullet hits school property for any reason, regardless of the number of victims, time of day, or day of week.

\subsubsection{Legal and Judicial}
The Systematic Content Analysis of Litigation Events Open Knowledge Network (SCALES-OKN) dataset \cite{scales-okn-2021} comprises two underlying datasets: PACER, the official source for electronic federal judicial records, and the Federal Judicial Center’s (FJC) database of appointed federal judges. SCALES-OKN incorporates some of PACER’s docket reports (ten years of docket reports from Northern Illinois district courts from 2007 to 2016, and docket reports from every district court in 2016) and judge metadata (birthdate, gender, race/ethnicity, history of appointments, appointing parties, education, and professional career).

\subsubsection{Healthcare}
The Medical Information Mart for Intensive Care (MIMIC-IV-ED) database \cite{johnson-2023} contains critical care data for over 40,000 patients (with patient identifiers removed according to the Health Insurance Portability and Accountability Act (HIPAA) Safe Harbor provision) admitted to intensive care units at the Beth Israel Deaconess Medical Center (BIDMC). MIMIC-IV-ED adopts a modular approach to data organization, highlighting data provenance and facilitating both individual and combined use of disparate data sources.

\subsubsection{Business}
The Yelp Open Dataset \cite{yelp-no-date} is a subset of Yelp’s businesses, reviews, and user data for use in personal, educational, and academic purposes. This dataset was collected by Yelp, and draws upon 5,996,996 reviews, 188,593 businesses, and 280,992 pictures from the Yelp platform.

\subsubsection{Environmental Sustainability}
Within this domain, we bring together two datasets. The first is the Air Quality Data Collected at Outdoor Monitors Across the US \cite{united-states-environmental-protection-agency-2015}. It contains an annual summary of Air Quality Index (AQI) values (an indicator of overall air quality taking into account all of the criteria air pollutants measured within a geographic area) for counties or core based statistical areas (CBSA). The summary values, which include both qualitative measures (days of the year having “good” air quality, for example) and descriptive statistics (median AQI value, for example), may vary in availability by area on account of many areas having monitoring stations for some, but not all, of the pollutants.
 
The second is the Spatial Wildfire Occurrence data for the United States \cite{USwildfire-2022}. This dataset contains spatial information about wildfires that occurred in the United States from 1992 to 2020 drawn from the records of federal, state, and local fire organizations.  It includes 2.3 million geo-referenced wildfire records, representing a total of 180 million acres burned during the 29-year period, as well as identifiers necessary to link the point-based, final-fire-reporting information to published large-fire-perimeter and operational-situation-reporting datasets.

\subsubsection{Urban Housing}
The Zillow Observed Rent Index \cite{zillow-group-inc-2023} is a rental price index designed to accurately represent the entire rental housing market, not just the properties currently listed for rent. It achieves this by considering the entire rental housing stock. This index is expressed in dollars and is calculated by determining the average rental prices falling within the 40th to 60th percentile range for all homes and apartments in a specific region. This calculation is performed at different geographical levels such as national, metropolitan, county, city, and zip code, as long as there is enough data available. To maintain accuracy, these calculations are weighted to account for the distribution of rental properties within the area.

\subsubsection{Socioeconomic}
Within this domain, we bring together datasets from two primary sources. The first is the Personal Income by County, Metro, and Other Areas report from the U.S. Bureau of Economic Analysis captures the (not seasonally adjusted) personal income of a country, metro, or other area (the income that is received by, or on behalf of, all the individuals who live in said area; estimates of personal income are presented by the place of residence of the income recipients) \cite{us-bureau-of-economic-analysis-personal-income-2022, us-bureau-of-labor-statistics-unemployed-2023}.

The second is The U.S. Census Bureau’s Small Area Income and Poverty Estimates (SAIPE) program provides annual estimates of income and poverty statistics for all school districts, counties, and states \cite{us-census-bureau-age-17-2022, us-census-bureau-median-household-2022, us-census-bureau-all-age-2022}. The program produces annual estimates of, among other measures, the number of children under age 18 in poverty.

\subsection{Generating Information Spaces}
\label{sec:gen_info_spaces}

The definition of the domain labelings in tandem with the analytics taxonomy makes possible the generation of an \emph{information space}: the set of all meaningful information that this data can convey given the available analysis.

\subsubsection{Plan Generation} 

\begin{table*}[ht!]
\small
\centering
\begin{tabular}{p{0.15\linewidth}p{0.70\linewidth}}
\hline
\textbf{Domain} & \textbf{Questions} \\
\hline
\multirow{2}{\linewidth}{Environmental Sustainability}
& What is the max air quality index grouped by year for state of Washington? \\
& What is the average fire size grouped by year for state of California? \\
\hline
\multirow{2}{*}{Healthcare}
& What is the disease for stay id of 31945330? \\
& What is the count of stay id for subject id of 10023239? \\
\hline
\multirow{3}{*}{Urban Housing}
& For date sorted in descending order and limited to the top results, what is the average rent for region name of United
States? \\
& What is the average rent for region name of San Francisco, CA? \\
\hline
\multirow{3}{*}{Criminal Justice}
& What is the count of unique incident id for weapon type containing "handgun"? \\
& Is the count of unique incident id for weapon type containing "handgun" greater than count of unique incident id for
weapon type containing "rifle"? \\
\hline
\multirow{3}{*}{Education}
& What is the correlation between total per pupil expenditure and percentage of students with an sat-math scores that exceed
standards for county of DuPage County? \\
& What is the average student enrollment for county of Adams and school type of HIGH SCHOOL? \\
\hline
\multirow{2}{*}{Legal and Judicial}
& What is the average case duration grouped by case type? \\
& What is the average case duration grouped by year for name of colleen kollar-kotelly? \\
\hline
\multirow{2}{*}{Socioeconomic}
& What is the average personal income grouped by year? \\
& What is the estimated people below 17 in poverty for county of Wayne County? \\
\hline
\multirow{2}{*}{Business}
& What is the average star rating for business of Lou Malnalti's? \\
& What is the count of business for city of Philadelphia? \\
\hline

\end{tabular}
\caption{A suite of example questions which can be surfaced via the plan generator. These represent a small subset of the useful information that can be derived for each of the domain labelings given the operations defined in the analytics taxonomy. Each question corresponds to an analytic plan which can be executed with the engine. In this way, meaningful information can be easily derived from any given dataset by specifying the lightweight domain labeling.}
\label{tab:example_questions}
\end{table*}

We produce analytic plan templates corresponding to these information goals which have fillable slots, allowing for executable plans to be generated and executed within the target domain by filling in these slots. Slot types include entity, attribute, and analysis operation slots.

To generate the space of possible plans, all combinations of the objects in the domain labeling that match a slot type, are used for filling that slot. For example, for a labeling defining a legal domain, the entity slot may be filled by a \textit{Judge} or a \textit{Case} entity. An attribute slot may be filled by a \textit{name} or \textit{age} attribute (if the entity slot has been filled by the \textit{Judge} entity); or \textit{case name}, \textit{duration}, or \textit{year} (if the entity slot has been filled by the \textit{Case} entity). In this way, the attribute slot is constrained by the entity it is being drawn from. Attribute slots are also constrained according to their attribute type. For example, a \textit{Metric} attribute slot may only be filled with attributes of this type (e.g., \textit{salary}, \textit{duration}, etc.), and a \textit{Categorical} attribute slot may only be filled by categorical attributes (e.g., \textit{case type}, \textit{year}, etc.). This separation of types is particularly useful for plan templates that incorporate grouping aggregations, as the categorical attribute will generally be the attribute to group on, whereas the metric will generally be the attribute to aggregate. Finally, an analysis operation slot may be filled by any analysis operation whose inputs match the plan structure at hand. In many cases, this is an aggregation (e.g., sum, max, min, average, etc.). The plans generated by filling these parameter slots with all possible combinations of valid arguments comprises the portion of the information space representing information goals that ask questions about the data without referring to a specific instances of \emph{entities} in the data (e.g., "What is the average case duration grouped by year?").

Additionally, the information space includes plans that answer questions about specific instances of \emph{entities} in the domain labeling, rather than questions about aggregations over all \emph{entities} in the data (as the information goals described above do). For example, "What is the average case duration grouped by year for judge name coleen kollar-kotelly?" This is done by adding a filler slot to the plan that limits it to a specific entity instance. Plans of this type are generated by filling this slot with each possible unique instance of this entity in the data. For example, for a \texttt{judge = [name]} filter, a different plan is generated by substituting \texttt{[name]} with each of the possible judge names in the data. Together, by filling these templates with possible combinations of values from the domain labeling, analytic taxonomy, and database, a vast, yet tractable space of useful analytic plans is surfaced to the user.

\subsubsection{Question Generation} 
\label{sec:question_gen}

Since graph-based plan representations are not the most intuitive way for lay-people to understand what kind of information can be derived from the data, we build a template-based language generator which can produce a corresponding question for each of the analytic plans. To generate a question, the plan graph is traversed in reverse order starting from the terminal \textit{return} operation node. The \emph{attributes} to be included are determined by examining the \textit{collect} operation referenced by the \textit{return} operation references. For example, the \textit{return} operation in Figure \ref{fig:plan_example} (step \texttt{|10|}) references \textit{collect} (step \texttt{|9|}), which references retrieved attributes \textit{year} and \textit{average amount grouped by year} (steps \texttt{|5|} and \texttt{|7|} respectively). Filters are expressed by examining filter operations referenced by the \textit{return} statement and mentioning them using a predefined "nicename" (e.g., "of" for \textit{exact}, "greater than" for \textit{greaterthan}, etc.). Similar language patterns are used for other operations. So, for example, the question generated from this plan is "What is the average amount of carbon emissions grouped by year in ascending order for country of United States of America?"

The analytic taxonomy and domain labeling specify language templates for the analysis operations and nicenames for the \emph{entity} \emph{attributes}, respectively. The analytic language templates are filled in using the \emph{attribute} nicenames in order to produce simple language statements that can be used to express the meaning of each step of the analytic plan. These simple language statements can be chained as necessary (for nodes with ancestors) and joined together (for plans that return multiple outputs). The result of this is a natural language question describing a corresponding analytic plan. Figure \ref{fig:ontology_implementation} depicts how the plan generator leverages the domain labeling, analytics taxonomy, and analytics engine to generate the space of possible questions. Examples of questions for each of the 8 domains that were generated with this method can be seen in Table \ref{tab:example_questions}.

These sample questions demonstrate examples of possible questions that get surfaced for a particular domain labeling, and represent the space of meaningful information that a user may be interested in deriving.

\section{Conclusion}
In this paper, we introduce an analytics taxonomy that encodes knowledge of analytic operations and how they map onto data in order to replicate key pieces of knowledge that data scientists use to derive meaningful information from data. To complement this knowledge representation, we showcased a system for surfacing quality information from datasets in a wide range of domains. In the future, we will seek to encode more data science knowledge by identifying new \emph{attribute} types and increasingly granular categories of types. For instance, \emph{Metrics} could be further broken down into diagnostic and ranking metrics to distinguish between measures that can be used for competitive comparisons (e.g. sales performance) versus those that should not (e.g. heart rate of patients in a hospital). We will also be exploring ways in which the information derived by the analytics engine can be better communicated (e.g. via documents and visualizations). This work lays the foundation for further work towards automating data science and providing systems that allow all people, regardless of their technical background, to derive meaningful insights from their data.

\bibliography{aaai24}

\appendix

\section{Taxonomy of Analytic Operations}
\label{app:operations}

A listing of the operations currently available within the analytics taxonomy can be seen in Table \ref{tab:ontology_operations}. This taxonomy was designed with extensibility in mind, and currently implements most major SQL operations.

\begin{table*}[ht!]
\small
\centering
\begin{tabular}{p{0.15\linewidth}p{0.12\linewidth}p{0.05\linewidth}p{0.25\linewidth}p{0.05\linewidth}p{0.25\linewidth}}
\textbf{Operation} & \textbf{Operation Type}& \textbf{Input Arity} & \textbf{Input Attribute Types}& \textbf{Output Arity} &\textbf{Output Attribute Types} \\
\hline

\multirow{2}{*}{Average}
& \multirow{2}{*}{Aggregation} & $1$ & [Arithmetic, Metric] & $1$ & [Arithmetic, Metric]  \\
&  & $\leq1$ & [Grouping] &  \\
\hline

\multirow{2}{*}{Correlation}
& \multirow{2}{*}{Aggregation} & $2$ & [Arithmetic, Metric, Datetime] & $1$ & [Arithmetic, Metric, Datetime]  \\
&  & $\leq1$ & [Grouping] &  \\
\hline

\multirow{2}{*}{Count}
& \multirow{2}{*}{Aggregation} & $1$ & [Arithmetic, Metric] & $1$ & [Arithmetic, Metric]  \\
&  & $\leq1$ & [Grouping] &  \\
\hline

\multirow{2}{*}{Count Unique}
& \multirow{2}{*}{Aggregation} & $1$ & [Arithmetic, Metric] & $1$ & [Arithmetic, Metric]  \\
&  & $\leq1$ & [Grouping] &  \\
\hline

\multirow{2}{*}{Get One}
& \multirow{2}{*}{Aggregation} & $1$ & [Arithmetic, Metric, Datetime] & $1$ & [Arithmetic, Metric, Datetime]  \\
&  & $\leq1$ & [Grouping] &  \\
\hline

\multirow{2}{*}{Max}
& \multirow{2}{*}{Aggregation} & $1$ & [Arithmetic, Metric, Datetime] & $1$ & [Arithmetic, Metric, Datetime]  \\
&  & $\leq1$ & [Grouping] &  \\
\hline

\multirow{2}{*}{Median}
& \multirow{2}{*}{Aggregation} & $1$ & [Arithmetic, Metric, Datetime] & $1$ & [Arithmetic, Metric, Datetime]  \\
&  & $\leq1$ & [Grouping] &  \\
\hline

\multirow{2}{*}{Min}
& \multirow{2}{*}{Aggregation} & $1$ & [Arithmetic, Metric, Datetime] & $1$ & [Arithmetic, Metric, Datetime]  \\
&  & $\leq1$ & [Grouping] &  \\
\hline

\multirow{2}{*}{Standard Deviation}
& \multirow{2}{*}{Aggregation} & $1$ & [Arithmetic, Metric] & $1$ & [Arithmetic, Metric]  \\
&  & $\leq1$ & [Grouping] &  \\
\hline

\multirow{2}{*}{String Aggregation}
& \multirow{2}{*}{Aggregation} & $1$ & [Arithmetic, Metric, Datetime] & $1$ & [Arithmetic, Metric, Datetime]  \\
&  & $\leq1$ & [Grouping] &  \\
\hline

\multirow{2}{*}{Sum}
& \multirow{2}{*}{Aggregation} & $1$ & [Arithmetic] & $1$ & [Arithmetic, Metric]  \\
&  & $\leq1$ & [Grouping] &  \\
\hline

And & Boolean & $\geq1$ & [Filter] & $1$ & [Filter] \\
\hline

\multirow{2}{*}{Contains}
& \multirow{2}{*}{Boolean} & $1$ & [Attribute] & $1$ & [Filter] \\
&  & $1$ & [Metric] &  \\
\hline

Exact & Boolean & $2$ & [Arithmetic, Metric, Categorical, String, Datetime, Identifier] & $1$ & [Filter] \\
\hline

Greater Than & Boolean & $2$ & [Arithmetic, Metric, Datetime] & $1$ & [Filter] \\
\hline

Greater Than Equal & Boolean & $2$ & [Arithmetic, Metric, Datetime] & $1$ & [Filter] \\
\hline

Less Than & Boolean & $2$ & [Arithmetic, Metric, Datetime] & $1$ & [Filter] \\
\hline

Greater Than Equal & Boolean & $2$ & [Arithmetic, Metric, Datetime] & $1$ & [Filter] \\
\hline

Not & Boolean & $1$ & [Filter] & $1$ & [Filter] \\
\hline

Or & Boolean & $\geq1$ & [Filter] & $1$ & [Filter] \\
\hline

Add & Arithmetic & $\geq2$ & [Arithmetic, Metric] & $1$ & [Arithmetic, Metric, Datetime] \\
\hline

Divide & Arithmetic & $\geq2$ & [Arithmetic, Metric] & $1$ & [Arithmetic, Metric, Datetime] \\
\hline

Multiply & Arithmetic & $\geq2$ & [Arithmetic, Metric] & $1$ & [Arithmetic, Metric, Datetime] \\
\hline

Percent Change & Arithmetic & $2$ & [Arithmetic, Metric] & $1$ & [Arithmetic, Metric] \\
\hline

Square Root & Arithmetic & $1$ & [Arithmetic, Metric, Datetime] & $1$ & [Arithmetic, Metric, Datetime] \\
\hline

Subtract & Arithmetic & $\geq2$ & [Arithmetic, Metric, Datetime] & $1$ & [Arithmetic, Metric, Datetime] \\
\hline

Collect & Data Operation & $\geq1$ & [Attribute] & $1$ & [AttributeCollection] \\
\hline

Groupby & Data Operation & $\geq1$ & [Categorical, Datetime] & $1$ & [Grouping] \\
\hline

Limit & Data Operation & $1$ & [Attribute] & $1$ & [Limit] \\
\hline

\multirow{4}{*}{Return}
& \multirow{4}{*}{Data Operation} & $1$ & [AttributeCollection] & $1$ & [Entity]  \\
&  & $\leq1$ & [Filter] &  \\
&  & $\leq1$ & [Sort] &  \\
&  & $\leq1$ & [Limit] &  \\
\hline

\multirow{2}{*}{Sort}
& \multirow{2}{*}{Data Operation} & $\geq1$ & [Attribute] & $1$ & [Sort] \\
&  & $1$ & [String] &  \\
\hline

\multirow{2}{*}{Retrieve Attribute}
& \multirow{2}{*}{Retrieval} & $1$ & [Entity] & $1$ & [Attribute] \\
&  & $1$ & [String] &  \\
\hline

Retrieve Entity & Retrieval & $1$ & [String] & $1$ & [Entity] \\
\hline

\end{tabular}
\caption{The set of all operations implemented as part of the analytics taxonomy. This includes operations used for retrieval, analysis, filtering, and data transformations.}
\label{tab:ontology_operations}
\end{table*}

\definecolor{codegreen}{rgb}{0,0.6,0}
\definecolor{codegray}{rgb}{0.5,0.5,0.5}
\definecolor{codepurple}{rgb}{0.58,0,0.82}
\definecolor{backcolour}{rgb}{0.95,0.95,0.92}

\lstdefinestyle{mystyle}{
    backgroundcolor=\color{backcolour},
    commentstyle=\color{codegreen},
    keywordstyle=\color{magenta},
    numberstyle=\tiny\color{codegray},
    stringstyle=\color{codepurple},
    basicstyle=\ttfamily\footnotesize,
    breakatwhitespace=false,
    breaklines=true,
    captionpos=b,
    keepspaces=true,
    numbers=left,
    numbersep=5pt,
    showspaces=false,
    showstringspaces=false,
    showtabs=false,
    tabsize=2
}

\lstset{style=mystyle}

\section{Examples of Analytic Operation Definitions}
\label{app:analytic_operations}
Below are examples of definitions of three of the analytic operations used by the system. Listing \ref{lst:avg_op} is a definition for the \emph{Average} aggregation operation, Listing \ref{lst:gte_op} is a definition for the \emph{Greater Than or Equal} data operation, and Listing \ref{lst:mult_op} is a definition for the \emph{Multiply} arithmetic operation.

The definitions comprise four components: the name, input argument types, output argument types, and language template. The name is the formal identifier that gets referenced when using the operation in the plan definition. The input argument types are used to restrict the types of values accepted as parameters to the operation, while the output types define the type of value that is output by the operation. The language template is used to aid in generating a question that corresponds to a plan where the operation is used.

\vspace{0.5in}

\begin{lstlisting}[language=Python, caption=Average Operation Definition, label={lst:avg_op}]
{
    "name": "average",
    "input_args": [
        {
            "arity": "1",
            "types": [
                "Arithmetic",
                "Metric"
            ]
        },
        {
            "arity": ">=1",
            "types": ["Grouping"]
        }
    ],
    "output_args": [
        {
            "arity": "1",
            "types": [
                "Arithmetic",
                "Metric"
            ]
        }
    ],
    "language_template": "average {0}"
}
\end{lstlisting}

\vspace{0.5in}

\begin{lstlisting}[language=Python, caption="Greater Than or Equal" Operation Definition, label={lst:gte_op}]
{
    "name": "greaterthan_eq",
    "input_args": [
        {
            "arity": "2",
            "types": [
                "Arithmetic",
                "Metric",
                "Datetime"
            ]
        }
    ],
    "output_args": [
        {
            "arity": "1",
            "types": ["Filter"]
        }
    ],
    "language_template": "{0} greater that or equal to {1}"
}
\end{lstlisting}

\vspace{2in}

\begin{lstlisting}[language=Python, caption=Multiply Operation Definition, label={lst:mult_op}]
{
    "name": "multiply",
    "input_args": [
        {
            "arity": ">=2",
            "types": [
                "Arithmetic",
                "Metric"
            ]
        }
    ],
    "output_args": [
        {
            "arity": "1",
            "types": [
                "Arithmetic",
                "Metric"
            ]
        }
    ],
    "language_template": "{0} multiplied by {1}"
}
\end{lstlisting}

\section{Domain Labeling Examples}
\label{app:domain_labelings}
While we plan to expand to additional data source types (e.g., knowledge graphs or documents), the domain labeling is currently only structured to support relational databases. Domain labelings are divided into three main sections: meta-data about the labeling ("id", "name", and "description"), data source schema information ("dataSource"), and the data abstraction layer definition ("dataAbstraction"). The data source schema information defines the tables and joins in the database. The data abstraction layer definition defines the \emph{entities} and \emph{relationships} between them. \emph{Entities} are defined in terms of one or more tables in the database, and their attributes are defined in terms of one or more columns from the \emph{entity}'s tables. \emph{Relationships} are defined in terms of one or more joins.

Note that construction of these domain labelings is semi-automated by having a user fill in the blanks (e.g. entities, attributes, relationships, joins, etc.) in a pre-annotated CSV file and then passing this to a script that converts this into the final JSON configuration format that actually gets ingested by the system upon startup. We are currently working to make this process more streamlined via an interactive interface, which is left as future work.

The JSON in Listing \ref{lst:aqi} depicts an example of a domain labeling configuration for the air quality database. Here, 2 \emph{entities} - AQI and Wildfire - are defined, each comprising a single corresponding table. A single \emph{relationship}, defined by the join that connects the underlying tables, links those two \emph{entities}.

The JSON in Listing \ref{lst:scales} example depicts a truncated domain labeling configuration for the SCALES-OKN database. Two \emph{entities} - Case and Judge - are defined. Here, the Case \emph{entity} comprises two database tables: "cases" (the primary table), and "case\_type". The latter table is used by the case\_type \emph{attribute} of the Case \emph{entity} to reference the "name" column from that table using the "casesTocase\_type" join. The \emph{relationship} linking the two \emph{entities} in this labeling, CaseToJudge, is a many-to-many \emph{relationship} (m2m) and as such comprises two joins: the one-to-many joining the "judges" table with the resolution table "judge\_on\_case" and the many-to-one joining the "judge\_on\_case" table with the "cases" table.

\onecolumn

\begin{lstlisting}[language=Python, caption=Air Quality Domain Labeling, label={lst:aqi}]
{
    "rid": "Toxic-3418-r4hu-4289-38jd93k29s",
    "name": "Air_Quality",
    "description": "Toxics and Air Quality (Jan 2000-Jun 2022)",
    "dataSource": {
        "type": "postgres",
        "connectionString": "<url_to_database>",
        "tables": [
            {"name": "aqi",         "primaryKey": {"id": "integer"}},
            {"name": "wildfire",    "primaryKey": {"id": "integer"}}
        ],
        "joins": [
            {"name": "wildfireToaqi", "from": "wildfire", "to": "aqi",
             "path": [["wildfire.county_state", "aqi.county_state", "string"]]}
        ]
    },
    "dataAbstraction": {
        "relationships": [
            {"name": "AQIToWildfire", "from": "AQI", "to": "Wildfire",
             "join": ["aqiTowildfire"], "relation": "o2o"}
        ],
        "entities": [
            {
                "name": "AQI",
                "nicename": ["Air Quality Index", "Air Quality Indices"],
                "table": "aqi",
                "id": "id",
                "idType": "integer",
                "attributes": {
                    "id": {
                        "nicename": ["ID", "IDs"],
                        "isa": "string",
                        "type": ["Identifier", "Categorical"],
                        "source": {"table": "aqi", "columns": ["id"]}
                    },
                    "state_name": {
                        "nicename": ["state", "states"],
                        "isa": "string",
                        "type": ["Categorical"],
                        "source": {"table": "aqi", "columns": ["state_name"]}
                    },
                    "county_name": {
                        "nicename": ["county name", "county names"],
                        "isa": "string",
                        "type": ["Categorical"],
                        "source": {"table": "aqi", "columns": ["county_name"]}
                    },
                    "year": {
                        "nicename": ["year", "years"],
                        "isa": "integer",
                        "type": ["Categorical"],
                        "source": {"table": "aqi", "columns": ["year"]}
                    },
                    "days_with_aqi": {
                        "nicename": ["days with air quality index",
                                     "days with air quality index"],
                        "isa": "integer",
                        "type": ["Arithmetic"],
                        "source": {"table": "aqi", "columns": ["days_with_aqi"]}
                    },
                    "good_days": {
                        "nicename": ["good days", "good days"],
                        "isa": "integer",
                        "type": ["Arithmetic"],
                        "source": {"table": "aqi", "columns": ["good_days"]}
                    },
                    "moderate_days": {
                        "nicename": ["moderate days", "moderate days"],
                        "isa": "integer",
                        "type": ["Arithmetic"],
                        "source": {"table": "aqi", "columns": ["moderate_days"]}
                    },
                    "unhealthy_for_sensitive_groups": {
                        "nicename": ["unhealthy for sensitive groups",
                                     "unhealth for sensitive groups"],
                        "isa": "integer",
                        "type": ["Arithmetic"],
                        "source": {"table": "aqi",
                                   "columns": ["unhealthy_for_sensitive_groups"]}
                    },
                    "unhealthy_days": {
                        "nicename": ["unhealthy days", "unhealthy days"],
                        "isa": "integer",
                        "type": ["Arithmetic"],
                        "source": {"table": "aqi", "columns": ["unhealthy_days"]}
                    },
                    "very_unhealthy_days": {
                        "nicename": ["very unhealthy days", "very unhealthy days"],
                        "isa": "integer",
                        "type": ["Arithmetic"],
                        "source": {"table": "aqi", "columns": ["very_unhealthy_days"]}
                    },
                    "hazardous_days": {
                        "nicename": ["hazardous days", "hazardous days"],
                        "isa": "integer",
                        "type": ["Arithmetic"],
                        "source": {"table": "aqi", "columns": ["hazardous_days"]}
                    },
                    "max_aqi": {
                        "nicename": ["max air quality index", "max air quality index"],
                        "isa": "integer",
                        "type": ["Arithmetic"],
                        "source": {"table": "aqi", "columns": ["max_aqi"]}
                    },
                    "90th_percentile_aqi": {
                        "nicename": ["90th percentile air quality index",
                                     "90th percentile air quality index"],
                        "isa": "integer",
                        "type": ["Arithmetic"],
                        "source": {"table": "aqi", "columns": ["90th_percentile_aqi"]}
                    },
                    "median_aqi": {
                        "nicename": ["median air quality index", "median air quality index"],
                        "isa": "integer",
                        "type": ["Arithmetic"],
                        "source": {"table": "aqi", "columns": ["median_aqi"]}
                    }
                }
            },
            {
                "name": "Wildfire",
                "nicename": ["Wildfire", "Wildfires"],
                "table": "wildfire",
                "id": "id",
                "idType": "integer",
                "attributes": {
                    "state_name": {
                        "nicename": ["state", "states"],
                        "isa": "string",
                        "type": ["Categorical"],
                        "source": {"table": "wildfire", "columns": ["state_name"]}
                    },
                    "county_name": {
                        "nicename": ["county name", "county names"],
                        "isa": "string",
                        "type": ["Categorical"],
                        "source": {"table": "wildfire", "columns": ["county_name"]}
                    },
                    "fire_size": {
                        "nicename": ["fire size", "fire sizes"],
                        "isa": "float",
                        "type": ["Arithmetic"],
                        "source": {"table": "wildfire", "columns": ["fire_size"]}
                    },
                    "fire_size_class": {
                        "nicename": ["fire size class", "fire size classes"],
                        "isa": "string",
                        "type": ["Categorical"],
                        "source": {"table": "wildfire", "columns": ["fire_size_class"]}
                    },
                    "discovery_time": {
                        "nicename": ["discovery time", "discovery times"],
                        "isa": "integer",
                        "type": ["Categorical"],
                        "source": {"table": "wildfire", "columns": ["discovery_time"]}
                    },
                    "discovery_date": {
                        "nicename": ["discovery date", "discovery dates"],
                        "isa": "date",
                        "type": ["Categorical"],
                        "source": {"table": "wildfire", "columns": ["discovery_date"]}
                    },
                    "contained_date": {
                        "nicename": ["contained date", "contained dates"],
                        "isa": "date",
                        "type": ["Categorical"],
                        "source": {"table": "wildfire", "columns": ["contained_date"]}
                    },
                    "fire_name": {
                        "nicename": ["fire name", "fire names"],
                        "isa": "string",
                        "type": ["Categorical"],
                        "source": {"table": "wildfire", "columns": ["fire_name"]}
                    },
                    "source_system_type": {
                        "nicename": ["source system type", "source system types"],
                        "isa": "string",
                        "type": ["Categorical"],
                        "source": {"table": "wildfire", "columns": ["source_system_type"]}
                    },
                    "year": {
                        "nicename": ["year", "years"],
                        "isa": "integer",
                        "type": ["Categorical"],
                        "source": {"table": "wildfire", "columns": ["year"]}
                    }
                }
            }
        ]
    }
}
\end{lstlisting}

\vspace{1in}

\begin{lstlisting}[language=Python, caption=Judicial Data Domain Labeling, label={lst:scales}]
{
    "rid": "SCALES-1467-4a02-8785-ec251e78d5be",
    "name": "scales",
    "description": "U.S. federal court records covering a variety of aspects of both civil and criminal proceedings, including parties, judges, attorneys, case metadata and docket entries.",
    "dataSource": {
        "type": "postgres",
        "connectionString": "<url_to_database>",
        "tables": [
            {"name": "cases", "primaryKey": {"ucid": "string"}},
            {"name": "case_type", "primaryKey": {"id": "integer"}},
            {"name": "judge_on_case", "primaryKey": {"id": "integer"}},
            {"name": "judges", "primaryKey": {"sjid": "string"}}
        ],
        "joins": [
            {
                "name": "casesTocase_type",
                "from": "cases",
                "to": "case_type",
                "path": [["cases.case_type_id", "case_type.id", "integer"]]
            },
            {
                "name": "judge_on_caseTojudges", 
                "from": "judge_on_case", 
                "to": "judges", 
                "path": [["judge_on_case.sjid", "judges.sjid", "string"]]
            },
            {
                "name": "judge_on_caseTocases", 
                "from": "judge_on_case", 
                "to": "cases", 
                "path": [["judge_on_case.ucid", "cases.ucid", "string"]]
            }
        ]
    },
    "dataAbstraction": {
        "relationships": [
            {
                "name": "CaseToJudge",
                "from": "Case",
                "to": "Judge",
                "join": ["judge_on_caseTojudges", "judge_on_caseTocases"],
                "relation": "m2m"
            }
        ],
        "entities": [
            {
                "name": "Case",
                "nicename": ["case", "cases"],
                "table": "cases",
                "id": "ucid",
                "idType": "string",
                "attributes": {
                    "case_id": {
                        "nicename": ["case ID", "case IDs"],
                        "isa": "string",
                        "type": ["Identifier"],
                        "source": {"table": "cases", "columns": ["case_id"]}
                    },
                    "ucid": {
                        "nicename": ["UCID", "UCIDs"],
                        "isa": "string",
                        "type": ["Identifier"],
                        "source": {"table": "cases", "columns": ["ucid"]}
                    },
                    "monetary_demand": {
                        "nicename": ["monetary demand", "monetary demands"],
                        "units": ["dollar", "dollars"],
                        "isa": "float",
                        "type": ["Arithmetic"],
                        "source": {"table": "cases", "columns": ["monetary_demand"]}
                    },
                    "billable_pages": {
                        "nicename": ["billable page", "billable pages"],
                        "isa": "integer",
                        "type": ["Arithmetic"],
                        "source": {"table": "cases", "columns": ["billable_pages"]}
                    },
                    "filing_date": {
                        "nicename": ["filing date", "filing dates"],
                        "isa": "date",
                        "type": ["Datetime"],
                        "source": {"table": "cases", "columns": ["filing_date"]}
                    },
                    "terminating_date": {
                        "nicename": ["terminating date", "terminating dates"],
                        "isa": "date",
                        "type": ["Datetime"],
                        "source": {"table": "cases", "columns": ["terminating_date"]}
                    },
                    "year": {
                        "nicename": ["year", "years"],
                        "isa": "integer",
                        "type": ["Categorical"],
                        "source": {"table": "cases", "columns": ["year"]}
                    },
                    "case_name": {
                        "nicename": ["case name", "case names"],
                        "isa": "string",
                        "type": ["Categorical"],
                        "source": {"table": "cases", "columns": ["case_name"]}
                    },
                    "jurty_demand": {
                        "nicename": ["jury demand", "jury demands"],
                        "isa": "string",
                        "type": ["Document"],
                        "source": {"table": "cases", "columns": ["jury_demand"]}
                    },
                    "cause": {
                        "nicename": ["cause", "causes"],
                        "isa": "string",
                        "type": ["Categorical"],
                        "source": {"table": "cases", "columns": ["cause"]}
                    },
                    "jurisdiction": {
                        "nicename": ["jurisdiction", "jurisdictions"],
                        "isa": "string",
                        "type": ["Categorical"],
                        "source": {"table": "cases", "columns": ["jurisdiction"]}
                    },
                    "case_flags": {
                        "nicename": ["case flag", "case flags"],
                        "isa": "string",
                        "type": ["Categorical"],
                        "source": {"table": "cases", "columns": ["case_flags"]}
                    },
                    "case_duration": {
                        "nicename": ["case duration", "case duration"],
                        "isa": "integer",
                        "type": ["Arithmetic"],
                        "source": {"table": "cases", "columns": ["case_duration"]}
                    },
                    "judge": {
                        "nicename": ["judge", "judges"],
                        "isa": "string",
                        "type": ["Categorical"],
                        "source": {"table": "cases", "columns": ["judge"]}
                    },
                    "referred_judges": {
                        "nicename": ["referred judge", "reffered judges"],
                        "isa": "string",
                        "type": ["Categorical"],
                        "source": {"table": "cases", "columns": ["reffered_judges"]}
                    },
                    "case_status": {
                        "nicename": ["case status", "case status"],
                        "isa": "string",
                        "type": ["Categorical"],
                        "source": {"table": "cases", "columns": ["case_status"]}
                    },
                    "cost": {
                        "nicename": ["cost", "costs"],
                        "units": ["dollar", "dollars"],
                        "isa": "float",
                        "type": ["Arithmetic"],
                        "source": {"table": "cases", "columns": ["cist"]}
                    },
                    "case_type": {
                        "nicename": ["case type", "case types"],
                        "isa": "string",
                        "type": ["Categorical"],
                        "source": {
                            "table": "case_type",
                            "columns": ["name"],
                            "joins": ["casesTocase_type"]
                        }
                    }
            },
            {
                "name": "Judge",
                "nicename": ["judge", "judges"],
                "table": "judges",
                "id": "id",
                "idType": "integer",
                "attributes": {
                    "sjid": {
                        "nicename": ["SJID", "SJIDs"],
                        "isa": "string",
                        "type": ["Identifier"],
                        "source": {"table": "judges", "columns": ["sjid"]}
                    },
                    "name": {
                        "nicename": ["name", "names"],
                        "isa": "string",
                        "type": ["Categorical"],
                        "source": {"table": "judges", "columns": ["name"]}
                    }
                }
            }
        ]
    }
}
\end{lstlisting}

\end{document}